\documentstyle[epsf]{l-aa}
\begin{document}

   \thesaurus{06         % A&A Section 6: Form. struct. and evolut. of stars
              (03.11.1;  % Cosmogony,
               16.06.1;  % Planets and satellites: general,
               19.06.1;  % Solar system: general,
               19.37.1;  % Stars: formation of,
               19.53.1;  % Stars: oscillations of,
               19.63.1)} % Stars: structure of.
%
%   \title{Radiation hydrodynamical modeling of the Circinus galaxy}
   \title{Circumnuclear obscuration of the Circinus galaxy
     by a starburst-radiation supported wall}

%   \subtitle{I. Overviewing the $\kappa$-mechanism}

   \author{K. Ohsuga \and M. Umemura 
%          \inst{1}
          }

%   \offprints{K. Ohsuga}

   \institute{Center for Computational Physics, University of Tsukuba,
              Tsukuba, Ibaraki 305-8577, Japan}

%   \date{Received September 15, 1989; accepted March 16, 1990}
   \date{Received 16 March 2000/ accepted 23 March 2001}
   \maketitle
        
\begin{abstract}
We consider the radiation-hydrodynamical 
formation of a dusty wall in the circumnuclear regions of 
the Circinus Seyfert 2 galaxy.
We focus on the radiative flux force 
due to the circumnuclear starburst, the nuclear starburst and
the active galactic nucleus (AGN),
and analyze the equilibrium configuration and stability of the dusty gas
in the circumnuclear regions.
The mass distributions used
to model the gravitational fields
are determined 
by the observed rotation velocities of the stars and of the gas.
Moreover, by using the simple stellar evolution in the circumnuclear starburst,
the bolometric luminosity of the starburst is estimated
as a function of time.
As a result, it is shown that 
the radiation force 
by the circumnuclear starburst 
does play an important role to build up 
a radiatively-supported dusty wall of $\la$ 100 pc,
which obscures the circumnuclear regions with $A_V$ of a few magnitudes.
It is also found that the 
age of a circumnuclear starburst when the circumnuclear regions are  
enshrouded by the dusty wall is constrained to be
$\la 10^8$ yr, which
is consistent with the age observationally inferred for 
the starburst, $4\, 10^7$ yr $-1.5\, 10^8$ yr.
%The Circinus galaxy may provide a precious example, in which
%the radiation force of a circumnuclear starburst could be related to 
%obscuring the circumnuclear regions.


\end{abstract}

%
%  14.Sep.'90: Demo-Vs.
%________________________________________________________________

\section{Introduction}
The structure of the obscuring material
around active galactic nuclei (AGNs)
is one of the most hotly debated issues.
Recently, 
it has been revealed that circumnuclear regions around AGNs
are embedded in obscuring material 
extending up to $\ga 100$ pc 
%(\cite{Rudy}; \cite{Miller}; \cite{Goodrich};
%\cite{McLeod}; \cite{Maio95}; \cite{MR95}; \cite{Malkan})
(Rudy et al. 1988; Miller et al. 1991; Goodrich 1995;
McLeod \& Rieke 1995; 
Maiolino et al. 1995; Maiolino \& Rieke 1995; 
Malkan et al. 1998),
although a dusty torus of on the parsec scale is also thought to 
surround AGNs according to the unified model (Antonucci 1993, for a review).
These facts may suggest that the obscuring material
around an AGN is distributed not only in nuclear regions, but also
in the circumnuclear regions on scales larger than some tens pc.
The Circinus galaxy which is a nearby Seyfert 2 galaxy with 
a circumnuclear starburst is one of the most suitable objects 
to study the structure of the obscuring material in circumnuclear regions,
since the galaxy has been studied on scales down to several
tens parsecs.
The $N_{\rm H}$ derived from X-ray observations imply
the visual extinction, $A_V$, to be $\approx 2000$ mag
in the nuclear region within a few pc (Matt et al. 1999; \cite{Risaliti})
and $A_V$ inferred by IR and optical observations 
is a few 10 mag
(Marconi et al. 1994; Alexander et al. 2000; Wilson et al. 2000).
It should be noted that 
since the regions of X-ray emission is considerably inner 
than those of IR emission,
each inferred $A_V$ may correspond to the spatially different 
obscuring material.
On the other hand,
$A_V$ in the circumnuclear regions (100 pc scale) of this galaxy
is also estimated to be a few mags as based on IR and optical observations
(Maiolino et al. 1998; Oliva et al. 1999; Wilson et al. 2000). 
%And an upper limit on the extinction towards the scattering region is 
%found to be 7.7 mag
%based on observations of the polarized broad H$\alpha$ as well as Br$\gamma$
%(Alexander et al. 2000).

Recently, Ohsuga \& Umemura (1999) have proposed that
the radiation force by a circumnuclear starburst is likely
to play an important role in the circumnuclear regions,
and it contributes to the obscuration 
through a large-scale obscuring wall which is radiatively supported 
by the circumnuclear starburst.
The goal of this paper is 
to study the distributions of dusty gas in the circumnuclear regions 
of the Circinus galaxy
from a radiation-hydrodynamical point of view.
With this goal, we carefully treat, based on the recent observational data,
the radiation sources and gravitational sources 
in the circumnuclear regions of the Circinus galaxy.
Then, by considering the radiation force exerted both 
by the circumnuclear starburst 
and by the nuclear starburst as well as by the AGN,
we analyze the equilibrium configuration and the stability of 
dusty gas in the circumnuclear regions.
The paper is organized as follows.
In Sect. 2, the radiation fields and gravitational fields are modeled
based on the recent observations. In Sect. 3, the formation of a radiatively
supported obscuring wall is presented.
Section 4 is devoted to conclusions.

\section{Radiation fields and gravitational fields}

\subsection{Radiation sources}
The intrinsic bolometric luminosity of the AGN component in the Circinus galaxy
is estimated to be $10^{10} L_\odot$ (Moorwood et al. 1996),
and H$\alpha$ images revealed 
that enhanced star formation occurs in nuclear regions
(Marconi et al. 1994; Elmouttie et al 1998a; Wilson et al. 2000).
However, the AGN itself of the Circinus galaxy is
thought to be surrounded by
the obscuring material with an $A_V$ of the order of a few tens 
to a thousand mags in the vicinity of the nucleus
thus reducing its effect on the surrounding regions
(Marconi et al. 1994; 
Matt et al. 1999; Alexander et al. 2000; Wilson et al. 2000).
But, the nuclear starburst
could exert radiation force on the dusty gas in circumnuclear regions.
Maiolino et al. (1998) estimated that 
the bolometric luminosity of the nuclear starburst 
is only 20\% or less of the AGN.
%whereas this estimation depends on the starburst model and 
%the extinction of nuclear starburst regions.
%the UV and optical radiation like an observed strong H$\alpha$ emission 
%from the nuclear regions (Elmouttie et al 1998a; Wilson et al. 2000)
On the other hand,
%In circumnuclear regions of the Circinus galaxy,
the presence of a starforming ring at a radius of $\sim 200$ pc 
is reported by Marconi et al. (1994), Elmouttie et al. (1998a), and
Wilson et al. (2000).
Based on the observed $K$-band luminosity and using
stellar population synthesis models,
Maiolino et al. (1998) have also argued that
the bolometric luminosity of the circumnuclear starburst
is $1.1\, 10^{10}L_\odot$ and its age 
is between $4\, 10^7$ yr and $1.5\, 10^8$ yr.
Therefore, we assume two components of radiation sources,
a nuclear starburst whose bolometric luminosity ($L_{\rm Nuc}$) is 
$2\, 10^{9}L_\odot$ and a starburst ring of 200 pc whose 
bolometric luminosity ($L_{\rm Ring}$) 
is $1.1\, 10^{10}L_\odot$ at the present day.

\subsection{Gravitational sources}
Based on the rotation velocities of stellar components or gaseous
components in circumnuclear regions of $\la 300$ pc
measured by Maiolino et al. (1998),
most of the mass is not concentrated into a pointlike object
but extended to a few hundred parsecs.
The dynamical mass within the starburst ring
is assessed to be at least $10^9M_\odot$
because the gaseous velocity at 200 pc is around 150 $\rm km\, s^{-1}$.
Moreover, it is reasonable to suppose that 
the mass distributions within the ring are uniform
since the velocity is roughly proportional to the radius.
An upper limit of the mass of a putative black hole 
is estimated as $4\, 10^6 M_\odot$ by Maiolino et al. (1998)
and 
Greenhill (2000) 
has detected a nuclear pointlike dark mass of $1.3\, 10^6 M_\odot$.
Elmouttie et al. (1998a) also reported that
the starburst ring 
has an inclination angle of $40^\circ \pm 10^\circ$
and a rotation velocity of $\sim$ 350 km $\rm s^{-1}$.
This implies that the dynamical mass within the ring
lies between $4\, 10^9 M_\odot$ and $9\, 10^9 M_\odot$.
(Elmouttie et al. 1998b estimated  
the dynamical mass of the nucleus to be less than 3.9 $\, 10^9 M_\odot$
by the observed rotation velocity of a molecular ring or disk with 
radius of about 300 pc.
Although this conclusion does not agree with the estimate
by the rotation velocity of the starburst ring, $\sim 4$-$9\, 10^9M_\odot$,
it is not definitely inconsistent 
because the radiative flux force is comparable to the gravity 
in the vicinity of the ring
and the net force which works on the molecular gas could be weaker than 
the intrinsic gravity by the dynamical mass.)
In addition, 
Jones et al. (1999) discovered that 
an H {\scriptsize I} 
ring or a disk of $\sim$ 1 kpc is rotating with a speed of 
at least 200 km s$^{-1}$, so that
the total dynamical mass within 1 kpc is around $\sim 9\, 10^9M_\odot$.
Therefore,
we consider the gravitational potential that is determined by
four components; the galactic bulge, the inner bulge, the starburst ring,
and the central black hole.
Here, we assume the galactic bulge 
to be a uniform sphere whose mass and radius are respectively
$M_{\rm GB}$ and 1 kpc, 
the inner bulge to be a uniform sphere whose mass and radius are
respectively $M_{\rm IB}$ and $R_{\rm IB}$,
and the mass of the central black hole to be $1.3\, 10^6M_\odot$.
Here, it is noted that 
the gravity of the black hole does not play important role
outside the central ten pc 
even if we adopt the upper limit $4\, 10^6M_\odot$ as the black hole mass. 
Taking account of the uncertainties in the modeling,
we assume the mass of the inner bulge within 200 pc 
of between $10^9 M_\odot$ and $9\, 10^9 M_\odot$,
and $R_{\rm IB}$ of $\ga 200$ pc 
based on the rotation curve given by Maiolino et al. (1998).
The difference between the total mass ($\sim 9\, 10^9M_\odot$) within 1 kpc 
and the sum of other components is
ascribed to the galactic bulge, $M_{\rm GB}$.
Although the mass distribution on the several hundred parsec scale is unknown,
it does not 
have much influence on the circumnuclear structure of the dusty gas 
within a few hundred parsecs.
The mass of the starburst ring should be determined by the subtraction
of the inner bulge mass from the mass inferred by the rotation
velocity at 200 pc.
Hence,
the mass of the starburst ring, $M_{\rm Ring}$, is presumed to be 
between a few $10^8M_\odot$ and a few $10^9M_\odot$.
%between $2.5\, 10^8M_\odot$ and $1.3\, 10^9M_\odot$,
%since there is an uncertainty on
%the age of $4\, 10^7$ yr $- 1.5\, 10^8$ yr.
Also,
we adopt a simple starburst model 
consistent with that of Maiolino et al. (1998)
in order to take account of the luminosity evolution
of the starburst ring,
where a Salpeter-type initial mass function (IMF) for a mass range of
[$1M_\odot$, $60M_\odot$],
the star formation rate in the starburst regions,
${\rm SFR} \propto \exp( -t/10^7 \rm yr)$,
the mass-luminosity relation, $(l_*/L_{\odot})=(m_*/M_{\odot})^{3.7}$, 
and
the mass-age relation, $\tau=1.1\,10^{10} {\rm yr}(m_*/M_{\odot})^{-2.7}$,
are employed.
%for the well determined luminosity of $1.1\, 10^{10}L_\odot$.
%The detail is presented in the next subsection. 

\section{Radiatively supported obscuring wall}
Based upon the model constructed above and
taking the effects by the optical depth into account,
we calculate the radiative flux force which is exerted 
on the dusty gas.
The radiative flux force exerted by the starburst ring 
at a point of $(r,z)$ in cylindrical coordinates is given by
\begin{equation}
  f^i_{\rm Ring} = \frac{\chi}{c} \int 
  \frac{\rho_{\rm Ring}}{4\pi l_{\rm Ring}^2}
  \frac{1-\exp(-\tau/\cos\theta_{\rm Ring})}
  {\tau/\cos\theta_{\rm Ring}} n^idV,
  \label{Ring}
\end{equation}
%\begin{equation}
% f^i_{\rm rad}=\frac{\chi}{c}\int \frac{\rho_{\rm Ring}}{4 \pi l^2}
%  n^i dV + \frac{\chi}{c} \frac{i L_{\rm AGN}}
%  {4\pi \left( r^2+z^2 \right)^{3/2}},
%\label{rad force}
%\end{equation}
where $i$ denotes $r$ or $z$,
$\chi$ is the mass extinction coefficient for dusty gas 
(Umemura et al. 1998), 
$c$ is the light speed,
$dV$ is an infinitesimal volume element of the starburst ring,
$l_{\rm Ring}$ is the distance from $(r,z)$ to this element,
$\theta_{\rm Ring}$ is the viewing angle from this element,
$\rho_{\rm Ring}$ is the luminosity density of the starburst ring 
($L_{\rm Ring}=\int \rho_{\rm Ring} dV$),
$\tau$ is the total optical depth of the wall,
and $n^i$ is a directional cosine (Ohsuga \& Umemura 2001).
Similarly, the radiative flux force by the nuclear starburst
is presented by
\begin{equation}
  f^i_{\rm Nuc} = 
  \frac{\chi}{c}
  \frac{i L_{\rm Nuc}}{4\pi \left( r^2+z^2 \right)^{3/2}}
  \frac{1-\exp(-\tau/\cos\theta_{\rm Nuc})}
  {\tau/\cos\theta_{\rm Nuc}},
  \label{Nuc}
\end{equation}
where $\theta_{\rm Nuc}$ is the viewing angle from the center.

Using Eqs. (\ref{Ring}) and (\ref{Nuc}),
the equilibrium between the radiation force and the gravity
is given by
\begin{equation}
  -\frac{{\rm d}\Phi}{{\rm d}z}=
  f_{\rm Ring}^z+f_{\rm Nuc}^z+f_{\rm grav}^z=0
\end{equation}
in the $z$ (vertical) directions
and 
\begin{equation}
  -\frac{{\rm d}\Phi}{{\rm d}r}=
  \frac{j^2}{r^3}+f_{\rm Ring}^r+f_{\rm Nuc}^r+f_{\rm grav}^r=0   
\label{eqr}
\end{equation}
in the $r$ (radial) directions,
where $\Phi$ is the effective potential,
$j$ is the specific angular momentum of dusty gas, and
$f_{\rm grav}^i$ is the gravitational force.
These equations are simultaneously solved
to obtain the equilibrium configuration,
by also taking into account of the conditions for the stability,
\begin{equation}
  \frac{{\rm d}^2 \Phi}{{\rm d}z^2} > 0,
  \label{zstable}
\end{equation}
and 
\begin{equation}
  \frac{{\rm d}^2 \Phi}{{\rm d}r^2} > 0.
  \label{rstable}
\end{equation}

\begin{figure}
\epsfxsize=8.8cm \epsfbox{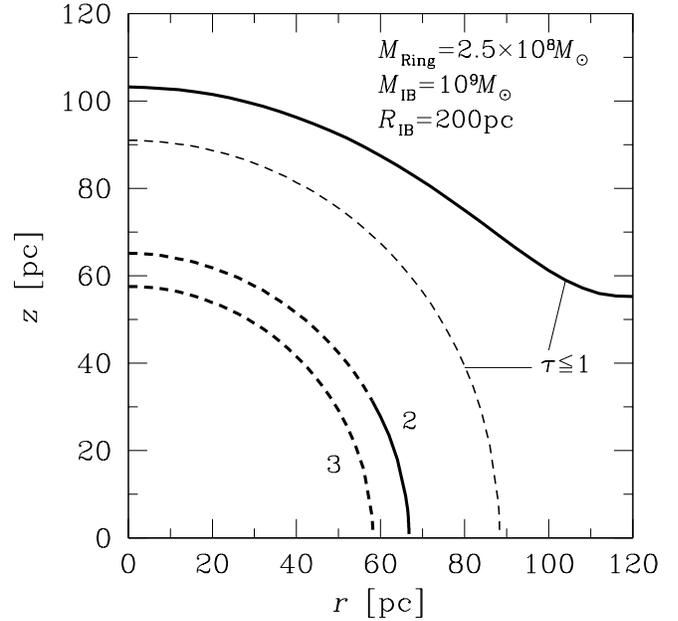}
%\picplace{2.0cm}
\caption[ ]
{
Equilibrium configuration of dusty gas between
the radiation force and gravity 
is shown in $r$-$z$ space for $\tau=3, 2$, and $\tau\leq 1$.
%two extreme cases of $M_{\rm IB}$.
Here, $M_{\rm Ring}=2.5\, 10^8 M_\odot$, $R_{\rm IB}=200$ pc,
%$m_{\rm low}=1.0M_\odot$, and $m_{\rm up}=60M_\odot$ are
and $M_{\rm IB}=10^9M_\odot$ are
assumed.
The thick and thin curves correspond to $t=4\, 10^7$ yr and 
$1.5\, 10^8$ yr, respectively.
In this model, the starburst age at
the present time is $4\, 10^7$ yr.
The solid curves represent stable branches,
while the dashed curves are stable in $z$ directions but 
radially nonequilibrium branches.
The solid curves show the final configuration of 
the stable obscuring wall.
The circumnuclear regions are
enshrouded by the obscuring wall of $\tau\sim$ a few at $4\times 10^7$ yr
}
\end{figure}
\begin{figure}
\epsfxsize=8.8cm \epsfbox{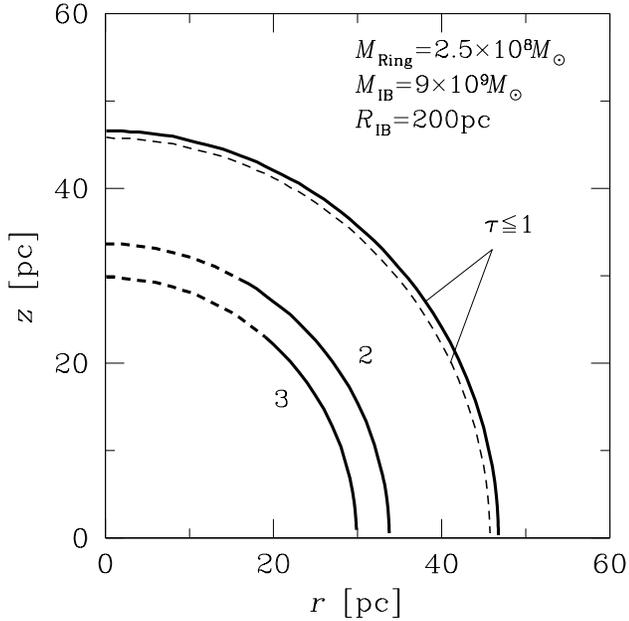}
%\picplace{2.0cm}
\caption[ ]
{
Same as Fig. 1, but for $M_{\rm IB}=9\, 10^9M_\odot$.
The circumnuclear regions are also 
enshrouded by the obscuring wall at $4\times 10^7$ yr.
Thus,
it is found that a stable wall forms which enshrouds
the circumnuclear regions at $4\, 10^7$ yr
regardless of uncertainties on gravitational fields
}
\end{figure}

In Fig. 1 and 2,
the resultant equilibrium branches are shown in the $r$-$z$ space
for $\tau=3, 2$ and $\tau\leq 1$.
Here, $M_{\rm Ring}=2.5\, 10^8 M_\odot$ and $R_{\rm IB}=200$ pc
%$m_{\rm low}=1.0M_\odot$, and 
%$m_{\rm up}=60M_\odot$
are assumed, and
the age of the starburst ring at
the present time is $4\, 10^7$ yr.
In both figures,
the thick curves show the branches at the present,
although the thin curves represent those at $1.5\, 10^8$ yr.
Also,
owing to the uncertainties in the observational estimation of the
inner bulge mass,
we consider two extreme cases of 
$M_{\rm IB}=10^9M_\odot$ in Fig. 1 and $M_{\rm IB}=9\, 10^9M_\odot$ in Fig. 2.
The former corresponds to the case that 
the inner bulge has the smallest density.
Although the inner bulge might extend to over 200 pc,
the equilibrium branches of $<$ 200 pc are not affected by
the size of the inner bulge itself and
depend solely on the mass within 200 pc.
The latter corresponds to the densest inner bulge.
In this case, $M_{\rm  GB}\sim 0$ and 
most of the mass is concentrated within 200 pc.
In Figs. 1 and 2,
all curves represent stable equilibrium branches in the $z$ directions.
Above the curves,
the $z$ component of the gravity, which works to lower the gas,
is stronger than that of radiation force,
while below the curves the radiation force lifts the gas towards 
the curves.
However, the stable equilibrium states in the $z$ directions do not always 
show the stable configuration of the obscuring wall.
To find it,
%totally stable configuration of the wall,
we need to check force balance (\ref{eqr}) and 
stable condition (\ref{rstable}) in the radial directions.
%It should be stressed that
In doing so we 
found that the force by the starburst ring plays an important role.
This can be more directly 
understood by the argument of azimuthal components of the forces,
%at a point on the wall.
where a unit vector in the azimuthal direction 
at a point of $(r,z)$ is $(-z, r)/(r^2+z^2)^{1/2}$.
Since the nuclear starburst, the bulge components,
and the black hole provide spherical forces,
they are not exerted on dusty gas in the azimuthal directions.
%where a directional cosine in the azimuthal direction 
%at a point of $(r,z)$ is $(-z, r)/(r^2+z^2)^{1/2}$.
%Since all branches are nearly spherical as shown in these figures,
%and the central source, the bulge components,
%and the black hole provide spherical force,
%these forces work perpendicular to the branches.
However, the centrifugal force and the force by the 
starburst ring have the azimuthal component.
Therefore, the centrifugal force can only balance with the ring force.
Since, as a result of the numerical integrations,
it is found that the azimuthal component of 
the radiation force by the ring is counterclockwise on these figures and
that of the gravity is clockwise in the same way 
as that of the centrifugal force,
the stable wall forms only when
the effective radiation force by the ring is stronger 
than the gravity by the ring at a point of branches.
If these components are out of balance, 
the dusty gas could not be in equilibrium there, and then 
the branches are of radial nonequilibrium.
In Fig. 1 and 2, 
the dashed curves are branches of radial nonequilibrium,
and the solid curves show the stable branches 
both in the $z$ (vertical) and in the $r$ (radial) directions.
Hence, these solid curves give the configuration of obscuring walls
which are expected to actually form.
%
%However, 
%the equilibrium states are not always stable in the radial directions.
%The effective potential in the radial directions 
%is not at local minimum on the dashed curves.
%In other words, the dashed curves are {\it radially} unstable branches.
%Resultantly, only the solid curves are both 
%{\it vertically} and {\it radially} stable branches.
%Hence, these curves give the configuration of obscuring wall
%which are expected to actually form.

Figures 1 and 2 show that a stable wall forms which enshrouds
the circumnuclear regions of several 10 pc $-100$ pc at $4\, 10^7$ yr
regardless of uncertainties on gravitational fields,
although the size is smaller for a denser inner bulge.
Then, the $A_V$ of the wall would be about a few mags
in spite of the uncertainties for the dust model, e.g. dust-to-gas mass ratio,
the size distribution, and the composition.
Near the rotation axis, 
the effective radiation force exerted on the wall
by the ring is weakened 
due to the large optical depth measured along the light ray
(see Eq. (\ref{Ring})),
so that the radial equilibrium is not achieved.
Thus, an opening forms near the axis.
On the other hand,
the wall turns to be radially nonequilibrium 
at $1.5\, 10^8$ yr even for $\tau\leq 1$,
since the bolometric luminosity of the ring decreases with time
due to stellar evolution.
(Here we do not take into account the time evolution 
of the bolometric luminosity
of the nuclear starburst, but
it does not have an influence on the force balance and stability 
in the $r$ directions of the dusty gas.)
\begin{figure}
\epsfxsize=8.8cm \epsfbox{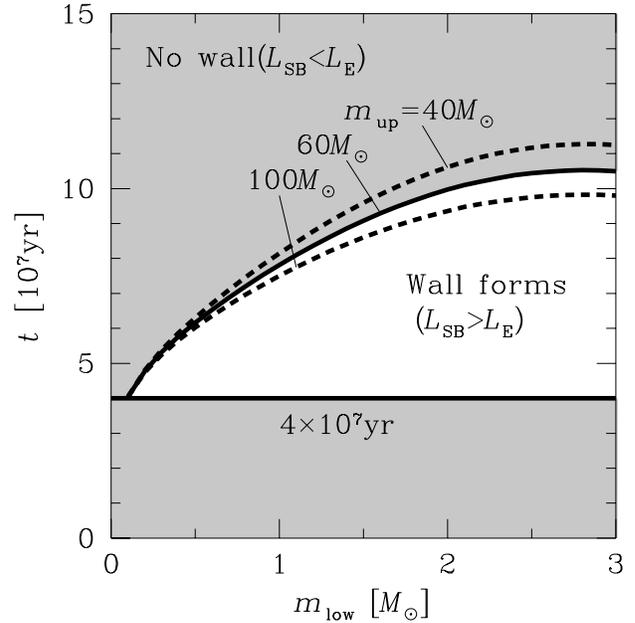}
%\picplace{2.0cm}
\caption[ ]
{
A condition for the age of a starburst ring on which a radiatively-supported 
stable wall can form is shown as a function 
of the low mass cutoff of the IMF, $m_{\rm low}$.
The range of the observationally inferred age, 
i.e., $4\times 10^7$ yr $-1.5\times 10^8$ yr, is considered.
Here, we assume the high mass cutoff of the IMF, $m_{\rm up}$, to be
$40 M_{\odot}$, $60 M_{\odot}$, or $100 M_{\odot}$.
Under the curves, the starburst ring is super-Eddington 
($L_{\rm Ring}>L_{E}=4\pi cGM_{\rm Ring}/\chi$), and then the radiation force
supports a stable wall which enshrouds the circumnuclear regions,
whereas no stable wall forms above the curves.
Hence, the radiation force plays a significant role in 
the Circinus galaxy, if the age of a circumnuclear starburst ring
lies in the white regions
}
\end{figure}
We have found that, also for other sets of parameters, the Eddington
ratio for the starburst ring itself gives a good criterion
for the formation of a wall which enshrouds the circumnuclear regions,
where Eddington luminosity is defined solely by the ring mass,
$L_{E}=4\pi cGM_{\rm Ring}/\chi$.
A condition for the starburst age for which a stable wall can form
is shown in Fig. 3, where we have considered
the range of the observationally inferred ages, 
i.e. $4\, 10^7$ yr $-1.5\, 10^8$ yr.
The condition depends upon the low mass cutoff of the IMF, $m_{\rm low}$,
although
almost regardless of a reasonable range of the upper mass limit, $m_{\rm up}$.
As a result, we find a range of the starburst ages 
compatible with the observation.
By taking into account the uncertainty on the IMF, it is 
concluded that the radiation force regulates 
the circumnuclear structure of the dusty gas in the Circinus galaxy 
if the age of the starburst is $\la 10^8$ yr.
In this case, 
the obscuration of the circumnuclear regions ($A_V$ of a few mags)
can be attributed to a radiatively-supported dusty wall.
Incorporating the present picture with the other observed features,
a schematic view of the Circinus galaxy is presented in Fig. 4.

\begin{figure}
\epsfxsize=8.8cm \epsfbox{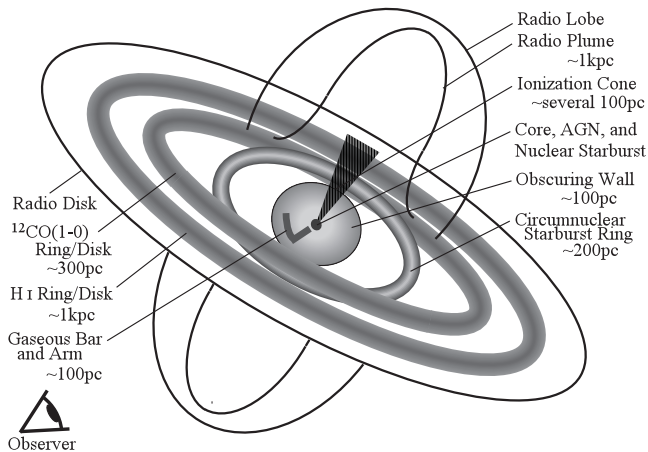}
%\picplace{2.0cm}
\caption[ ]
{
A schematic view of the nuclear regions of the Circinus galaxy
based upon the radiation-hydrodynamical modeling.
In the central regions of the Circinus galaxy,
there exist a Sy2 nucleus 
and a core which is observed by visible and near infrared
coronal lines (Oliva et al. 1994), as well as H${\alpha}$, and radio continuum 
emission (Elmouttie et al. 1998a).
Moreover, an inner 100pc bar and a spiral arm are discovered by 
Maiolino et al. (2000).
It is reported that the ionization cone 
detected in H$\alpha$ ($\sim$ 400 pc) 
and O[III] (500-900 pc) 
(Veilleux \& Bland-Hawthorn 1997; Elmouttie et al. 1998a;
Wilson et al. 2000)
lies in the direction of NW.
The obscuring wall is supported due to radiation force by 
the circumnuclear starburst ring 
as well as the nuclear starburst.
Hence, the AGN, the core, and the bottom of the ionization cone
would be obscured by the wall.
A disk is detected by observations of 
radio continuum (Elmouttie et al. 1995),
H {\scriptsize I} (Jones et al. 1999), and $^{12}$CO(1-0) (Elmouttie et al.1998b).
These observations also
revealed the ring-like features of $^{12}$CO(1-0) of $\sim$ 350 pc and 
HI of $\sim$ 1kpc, respectively.
The inclination angle of the starburst ring ($40^{\circ}\pm 10^{\circ}$) is very
different from the edge-on disk components.
In addition, radio lobes and plume are detected by Elmouttie et al. (1995).
The radio components are observed to be two-sided, whereas 
the ionization cone extends to the direction of NW only
}
\end{figure}

\section{Conclusions}
To study the distributions of dusty gas in the circumnuclear regions 
of the Circinus galaxy,
we have analyzed the equilibrium configurations and the stability of 
the dusty gas in the circumnuclear regions
by considering the radiation force exerted by the circumnuclear starburst ring
as well as by the central radiation source, which consists mainly 
of a nuclear starburst.
As a result, it has been found that,
within the observationally inferred age of a circumnuclear starburst ring
in the Circinus galaxy, the radiation force can work to build up a dusty wall
which enshrouds the circumnuclear regions.
The wall may contribute to the obscuration of the region inside the nuclear
100 pc in this galaxy.

The visual extinction $A_V$ of the wall could be a few mags.
Also, if dusty gas which was blown away by stronger starburst radiation
in earlier phases is afloat in the circumnuclear regions,
$A_V$ would ascend further.
Although the detail is not given before
multi-dimensional radiation hydrodynamics is treated
(will be performed in the future analysis),
$A_V$ of the circumnuclear regions of the Circinus galaxy would be 
a few or several mags. 
In any case,
the present picture provides a physical structure of the dusty gas
in the circumnuclear regions, which is consistent with the 
extended obscuring materials in the Circinus galaxy
(Maiolino et al. 1998; Oliva et al. 1999; 
Wilson et al. 2000).
%This estimate is consistent with 
%the steep extinction gradient between the nuclear and the extranuclear 
%regions, $\Delta A_V$, which is inferred to be from 2.5 to 3.0
%(Oliva et al. 1999).
%Although, $A_V$ of the dusty wall by itself may not be 
%enough to shift this AGN completely from type 1 to type 2,
%the wall may work cooperatively with 
%putative obscuring material of subparsec scale
%to regulate the type of the AGN.

\begin{acknowledgements}
We are grateful to the anonymous referee for valuable comments.
The calculations were carried out at Center for Computational Physics 
in University of Tsukuba. This work is 
supported in part by Research Fellowships of the Japan Society
for the Promotion of Science for Young Scientists, 6957 (KO)
and the Grants-in Aid of the
Ministry of Education, Science, Culture, and Sport, 09874055 (MU). 
\end{acknowledgements}


\begin{thebibliography}{}
\bibitem{}
%Alexander, D. M., Heisler, C. A., Young, S., Lumsden, S. L., Hough, J. H., Bailey, J. A. 2000, MNRAS 313, 815
Alexander, D. M., Heisler, C. A., Young, S., et al. 2000, MNRAS, 313, 815
\bibitem{}
Antonucci, R. 1993, ARA\&A, 31, 473
%\bibitem[Brandl et al. 1996]{Brandl}
%Brandl B., Sams B. J., Bertoldi F., Eckart A., Genzel R., Drapatz S., Hofmann R., L\"owe M., Quirrenbach A., 1996, ApJ 466, 254
%\bibitem[Charlot et al. 1993]{Charlot}
%Charlot S., Ferrari F., Mathews G. J., Silk J., 1993, ApJ 419, L57
%\bibitem[Doane \& Mathews 1993]{Doane}
%Doane J. S., Mathews W. G., 1993, ApJ 419, 573
%\bibitem[Doyon, Puxley, \& Joseph 1992]{Doyon}
%Doyon R., Puxley P. J., Joseph R. D., 1992, ApJ 397, 117
\bibitem[Elmouttie et al. 1995]{Elmouttie95}
Elmouttie, M., Haynes, R. F., Jones, K. L., et al. 1995, MNRAS, 275, L53
%Elmouttie M., Haynes R. F., Jones K. L., Ehle M., Beck R., Wielebinski R. 1995, MNRAS 275, L53
\bibitem[Elmouttie et al. 1998a]{Elmouttie98a}
Elmouttie, M., Koribalski, B., Gordon, S., et al. 1998a, MNRAS, 297, 49
%Elmouttie M., Koribalski B., Gordon S., Taylor K., Houghton S., Lavezzi T., Haynes R., Jones K. 1998a, MNRAS 297, 49
\bibitem[Elmouttie et al. 1998b]{Elmouttie98b}
Elmouttie, M, Krause, M., Haynes, R. F., \& Jones, K. L. 1998b, MNRAS, 300, 1119
\bibitem[Goodrich 1995]{Goodrich}
Goodrich, R. W. 1995, ApJ, 440, 141
%\bibitem[Heckman et al. 1989]{Heckman}
%Heckman T. M., Blitz L., Wilson A. S., Armus L., Miley G. K. 1989, ApJ 342, 735
%\bibitem[Hill et al. 1994]{Hill}
%Hill J. K., Isensee J. E., Cornett R. H., Bohlin R. C., O'Connell R. W., Roberts M. S., Smith A. M., Stecher, T. P. 1994, ApJ 425, 122
%\bibitem[Hunt et al. 1997]{Hunt}
%Hunt L. K., Malkan M. A., Salvati M., Mandolesi N., Palazzi E., Wade R. 1997, ApJS 108, 229
\bibitem{}
Greenhill, L. J. 2000 [astro-ph/0010277]
\bibitem[Jones et al. 1999]{Jones}
Jones, K. L., Koribalski, B. S., Elmouttie, M., \& Haynes, R. F. 1999, MNRAS, 302, 649
\bibitem[McLeod \& Rieke 1995]{McLeod}
McLeod, K. K., \& Rieke, G. H. 1995, ApJ, 441, 96
\bibitem[Maiolino et al. 1995]{Maio95}
Maiolino, R., Ruiz, M., Rieke, G. H., \& Keller, L. D. 1995, ApJ, 446, 561
\bibitem[Maiolino \& Rieke 1995]{MR95}
Maiolino, R., \& Rieke, G. H. 1995, ApJ, 454, 95
%\bibitem[Maiolino et al. 1997]{Maio97}
%Maiolino R., Ruiz M., Rieke G. H., Papadopoulos P. 1997, ApJ 485, 552
\bibitem[Maiolino et al. 1998]{Maio98}
Maiolino, R., Krabbe, A., Thatte, N., \& Genzel, R. 1998, ApJ, 493, 650
%\bibitem[Maiolino, Risaliti, \& Salvati 1999]{Maio99}
%Maiolino R., Risaliti G., Salvati M. 1999, A\&A 341, L35
\bibitem[Maiolino et al. 2000]{Maio00}
Maiolino, R., Alonso-Herrero, A., Anders, S., et al. 2000, ApJ, 531, 219
\bibitem[Malkan, Gorjian, \& Tam 1998]{Malkan}
Malkan, M. A., Gorjian, V., \& Tam, R. 1998, ApJS, 117, 25
\bibitem[Marconi et al. 1994]{Marconi}
Marconi, A., Moorwood, A. F. M., Origlia, L., \& Oliva, E. 1994, ESO Messenger, 78, 20
\bibitem{}
Matt, G., Guainazzi, M., Maiolino, R., et al. 1999, A\&A, 341, L39
\bibitem[Miller, Goodrich, \& Mathews 1991]{Miller}
Miller, J. S., Goodrich, R. W., \& Mathews, W. G. 1991, ApJ, 378, 47
%\bibitem{}
%Moorwood A. F. M., Glass, I. S. 1984, A\&A 135, 281
\bibitem[Moorwood et al. 1996]{Moorwood}
Moorwood, A. F. M., Lutz, D., Oliva, E., et al. 1996, A\&A, 315, L109
\bibitem[Ohsuga \& Umemura 1999]{Ohsuga}
Ohsuga, K., \& Umemura, M. 1999, ApJ, 521, L13
\bibitem{}
Ohsuga, K., \& Umemura, M. 2001, submitted
\bibitem[Oliva et al. 1994]{Oliva94}
Oliva, E., Salvati, M. Moorwood, A. F. M., \& Marconi, A. 1994, A\&A, 288, 457
%\bibitem{}
%Oliva, E., Origlia, L., Kotilainen, J. K., Moorwood, A. F. M. 1995, A\&A 301, 55
\bibitem[Oliva, Marconi, \& Moorwood]{Oliva00}
Oliva, E., Marconi, A., \& Moorwood, A. F. M. 1999, A\&A, 342, 87
%\bibitem[P\'erez-Olea \& Colina 1996]{Perez}
%P\'erez-Olea D. E., Colina L. 1996, ApJ 468, 191
%\bibitem{}
%Rieke, G. H., Lebofsky, M. J. 1985, ApJ 288, 618
\bibitem[Risaliti, Maiolino, \& Salvati 1999]{Risaliti}
  Risaliti, G., Maiolino, R., \& Salvati, M. 1999, ApJ, 522, 157
\bibitem[Rudy, Cohen, \& Ake 1988]{Rudy}
  Rudy, R. J., Cohen, R. D., \& Ake, T. B. 1988, ApJ, 332, 172
%\bibitem[Storchi-Bergmann, Schmitt, \& Fernandes 1999]{Storchi}
%Storchi-Bergmann T., Schmitt H. R., Fernandes R. C. 1999, 
%in IAU Symposium No. 194,
%Activity in Galaxies and Related Phenomena,
%ed. Y. Terzian, E. Khachikian, D. Weedman
%(San Francisco: ASP), 295
\bibitem[Umemura, Fukue, \& Mineshige 1998]{Umemura}
Umemura, M., Fukue, J., \& Mineshige, S. 1998, MNRAS, 299, 1123
\bibitem[Veilleux \& Bland-Hawthorn 1997]{Veilleux}
Veilleux, S., \& Bland-Hawthorn, J. 1997, ApJ, 479, L105
\bibitem{}
Wilson, A. S., Shopbell, P. L., Simpson, C., et al. 2000, AJ, 120, 1325
\end{thebibliography}
\end{document}